\documentclass[
nofootinbib,
amsmath,amssymb,
aps,
prd,preprint
]{revtex4-1}
\pdfoutput=1

\usepackage{setspace}
\usepackage{graphicx}
\usepackage{natbib}

\usepackage{hyperref}

\usepackage{ulem}
\usepackage{tikz}

\usepackage[utf8]{inputenc}
\begin{document}

\title{Simple single-field inflation models with arbitrarily small tensor/scalar ratio}
\author{Nina K. Stein}
\affiliation{Dept. of Physics, Univ. at Buffalo, SUNY}
\author{William H. Kinney}
\affiliation{Dept. of Physics, Univ. at Buffalo, SUNY}
\affiliation{Dept. of Physics, Indian Institute of Technology, Madras}
\date{December 27, 2022}

\begin{abstract}
   We construct a family of simple single-field inflation models consistent with Planck / BICEP Keck bounds which have a parametrically small tensor amplitude and no running of the scalar spectral index. The construction consists of a constant-roll hilltop inflaton potential with the end of inflation left as a free parameter induced by higher-order operators which become dominant late in inflation. This construction directly demonstrates that there is no lower bound on the tensor/scalar ratio for simple single-field inflation models. 
\end{abstract}

\maketitle

\section{Introduction}
Inflationary cosmology \cite{Starobinsky:1980te,Sato:1981ds,Sato:1980yn,Kazanas:1980tx,Guth:1980zm,Linde:1981mu,Albrecht:1982wi} remains a uniquely successful phenomenological framework for understanding the origin of the universe, making quantitative predictions which current data strongly support \cite{Spergel:2006hy,Alabidi:2006qa,Seljak:2006bg,Kinney:2006qm,Martin:2006rs}. Inflation relates the evolution of the universe to one or more scalar \textit{inflaton} fields, the properties of which dictate the dynamics of the period of rapidly accelerating expansion that terminates locally in a period of reheating, followed by radiation-dominated expansion. The specific form of the potential of the inflaton field or fields is unknown, but different choices of potential result in different values for cosmological parameters, which are distinguishable by observation \cite{Dodelson:1997hr,Kinney:1998md}. Recent data, in particular the Planck measurement of Cosmic Microwave Background (CMB) anisotropy and polarization \cite{Planck:2015sxf,Ade:2015xua,Aghanim:2015xee} and the BICEP/Keck measurement of CMB polarization \cite{Ade:2015fwj} now place strong constraints on the inflationary parameter space. Consequently, many previously viable  inflationary potentials, including some of the simplest and most theoretically attractive models, are in conflict  with these observations, in particular the upper bound on the tensor scalar ratio $r$. Furthermore, near-future measurements  could reduce this upper bound from $r\leq 10^{-2}$ to $r\leq 10^{-3}$. \cite{Easther:2021eje,CMB-S4:2016ple}

This raises the question: is there an upper bound on $r$ which would rule out all simple single-field inflation models? This question presupposes a definition of ``simple'', which is inherently subjective. Recent work has explored this question in the context of supersymmetry-inspired $\alpha$-attractor models of inflation \cite{Brooker:2017vyi,Kallosh:2019hzo}. In this paper, we will adopt a definition of ``simple'' that consists of a single scalar field, with canonical Lagrangian, and a potential that can be approximated during the epoch of inflation by a single leading-order operator. This has the advantage of being entirely generic, applying to any hilltop-type potential. Any arbitrary single-field potential can be represented by an effective operator expansion,
\begin{equation}
    V = V_0+\left.\frac{dV}{d\phi}\right\vert_{\phi=0}\phi+\frac{1}{2}\left.\frac{d^2V}{d\phi^2}\right\vert_{\phi=0}\phi^2+\frac{1}{6}\left.\frac{d^3V}{d\phi^3}\right\vert_{\phi=0}\phi^3+\cdots.
\end{equation}
Different choices of coefficients result in different inflationary dynamics and different predictions for observables, such as the tensor/scalar ratio $r$ and scalar spectral index $n_S$. This can be used to falsify particular inflationary models \cite{Dodelson:1997hr}.  For example, the simplest monomial potentials of the form
\begin{equation}
    V\left(\phi\right) \propto \phi^p,\quad p > 0,
\end{equation}
are ruled out, since they overproduce tensor perturbations, violating the upper bound set by the BICEP/Keck measurement \cite{BICEP:2021xfz}. Hilltop models, of the general form
\begin{equation}
    V\left(\phi\right) = V_0 - \lambda \left(\frac{\phi}{\mu}\right)^p
\end{equation}
fare better, where $\lambda$ is a dimensionless coupling constant, and $\mu$ is a mass scale determining the range of  validity of the effective expansion, with $\Delta \phi \sim \mu$ during inflation. The field excursion $\Delta\phi$ is directly related to the tensor/scalar ratio by the Lyth bound \cite{Lyth:1996im}, 
\begin{equation}
\Delta \phi \geq M_{\mathrm P} \sqrt{2 r},
\end{equation}
where $M_{\mathrm P}$ is the reduced Planck mass. ``Swampland'' conjectures, motivated by string theory \cite{Ooguri:2006in}, suggest that the field excursion is bounded in any effective field theory which can be completed in the ultraviolet, 
\begin{equation}
    \Delta\phi < M_{\mathrm P},
\end{equation}
with the consequence that there may be a corresponding upper bound on $r$ in viable inflationary models. A well-known example of a model which can accommodate arbitrarily small $r$ is one for which the mass term for the inflaton field $\phi$ is suppressed, for example by a shift symmetry, with the leading behavior
\begin{equation}
   V\left(\phi\right) = V_0\left[1 -  \left(\frac{\phi}{\mu}\right)^4 + \cdots \right]. 
\end{equation}
In this case, the scalar normalization depends only on the ratio of the height of the potential $V_0^{1/4}$ to the width $\mu$ \cite{Kinney:1995cc}, and taking $\mu < M_{\mathrm P}$ results in an upper bound on the tensor/scalar ratio $r$ of \cite{Kinney:2006qm}
\begin{equation}
    r < \frac{1}{4 N^3}.
\end{equation}
The scalar spectral index  $n_S$ in the limit $\mu \ll M_{\mathrm P}$ similarly depends only on the number of e-folds $N$ of inflation, 
\begin{equation}
    n_S = 1 - \frac{3}{N}. 
\end{equation}
The number of e-folds of inflation, $N$, depends on the reheat temperature, but is bounded by $N \leq 60$, which corresponds to an upper bound on $n_S$ of
\begin{equation}
    n_S < 0.95,
\end{equation}
placing it just outside the region allowed by the Planck measurement of the CMB, $n_S = 0.9649 \pm 0.0042$ \cite{Planck:2018jri}.  
\begin{figure}
\includegraphics[scale=0.30]{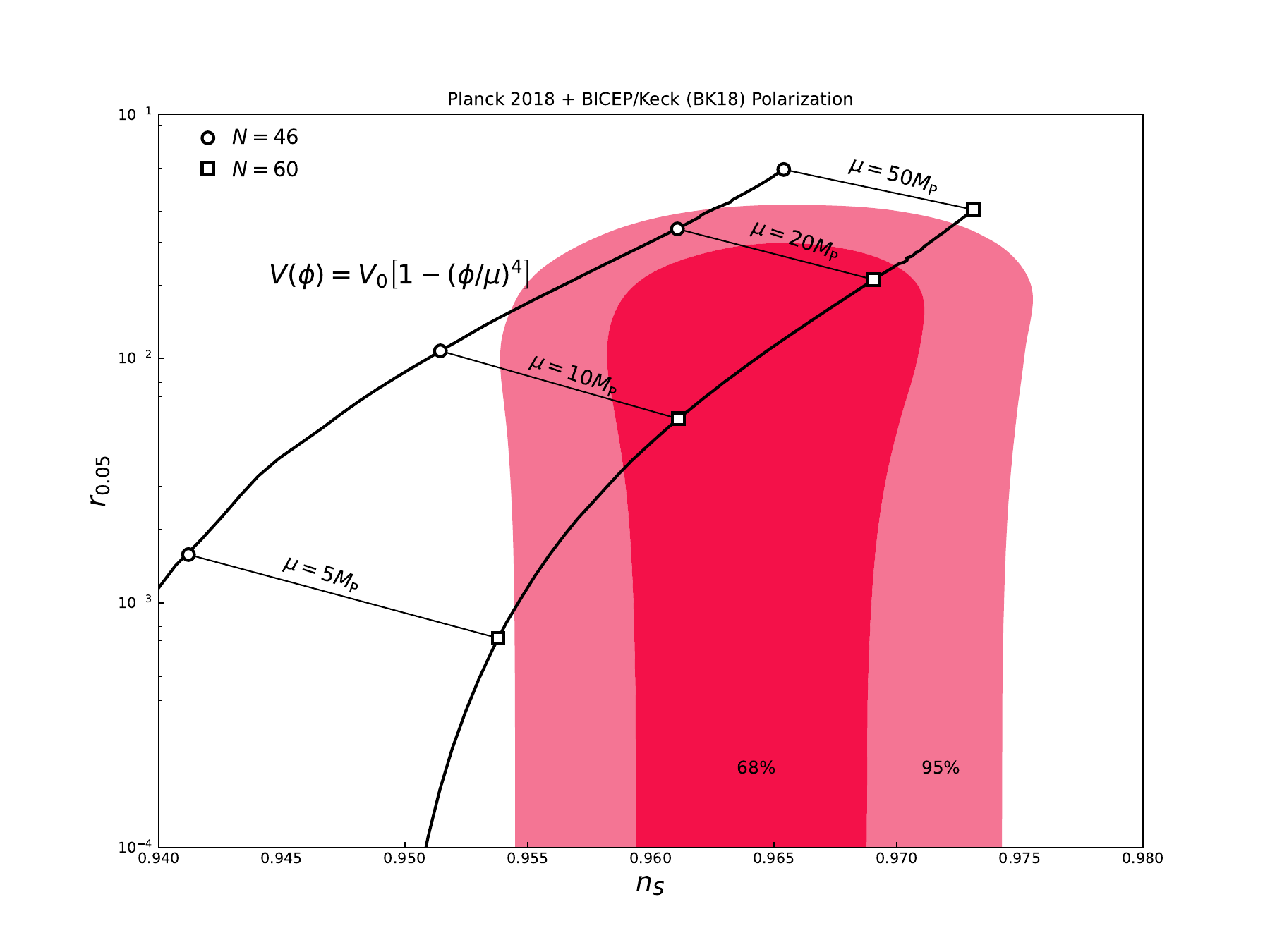}
\includegraphics[scale=0.30]{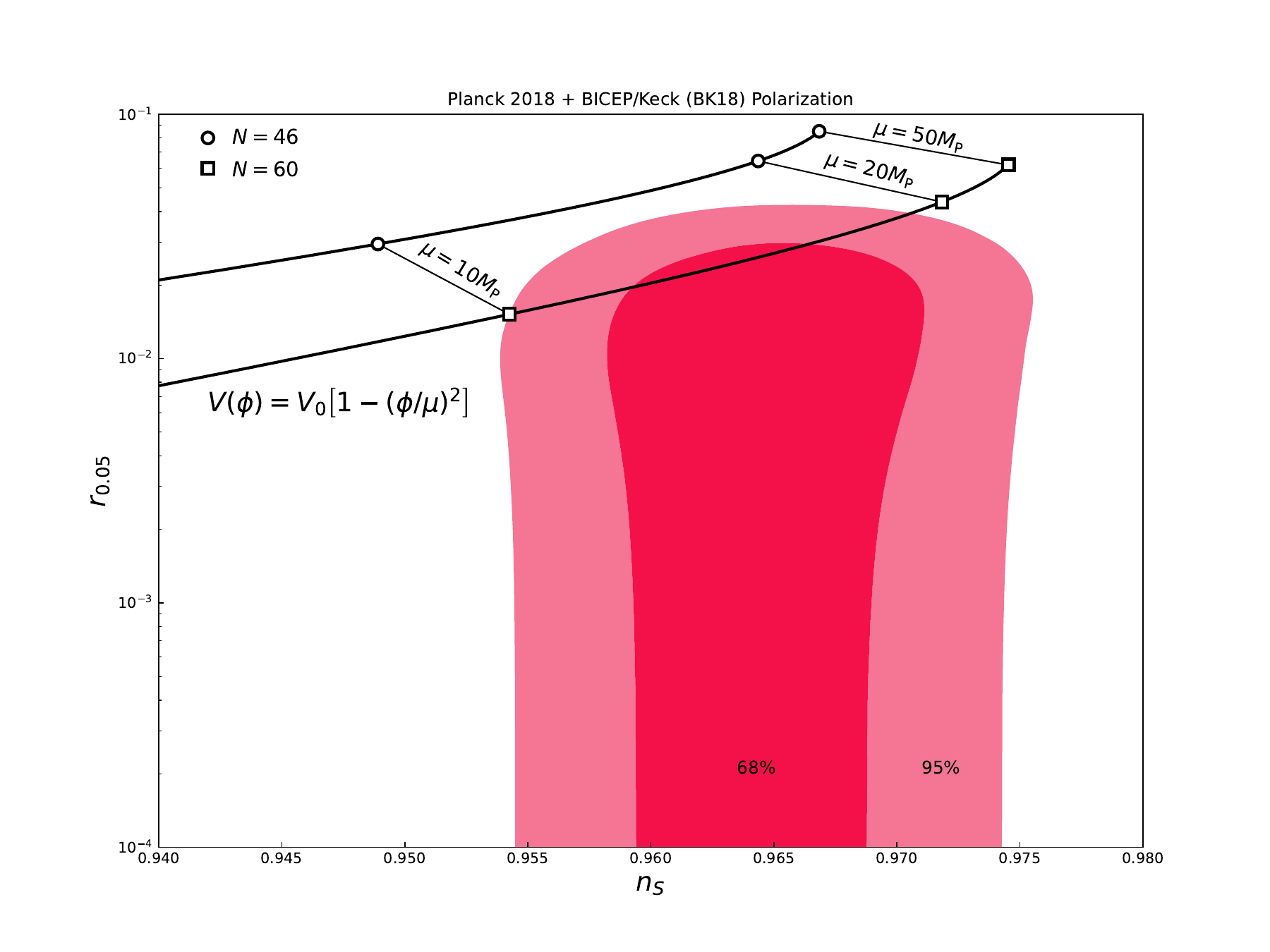}
\caption{Predictions of the quartic (top) and quadratic (bottom) potentials compared to the allowed region from Planck \cite{Planck:2018jri} and BICEP/Keck \cite{BICEP:2021xfz}. Shaded regions show the allowed parameter regions at 68\% and 95\% confidence, and plotted lines show the predictions for the potentials for various choices of mass parameter $\mu$. Of particular note is that the small-$r$ limit of the quartic potential is inconsistent with the Planck constraint on the spectral index $n_S$. }
\label{fig:Phi4}
\end{figure}
Conversely, for a potential whose leading-order behavior is quadratic,
\begin{equation}
   \label{eq:quadpotential}
   V\left(\phi\right) = V_0\left[1 -  \left(\frac{\phi}{\mu}\right)^2 + \cdots \right],
\end{equation}
the spectral index $n_S$ is \cite{Kinney:2006qm}
\begin{equation}
    \label{eq:quadnS}
    n_S = 1 - 4 \left(\frac{M_\mathrm{P}}{\mu}\right)^2.
\end{equation}
The tensor/scalar ratio depends on the number of e-folds $N$ and the spectral index $n_S$ as
\begin{equation}
    \label{eq:quadr}
    r = 8 \left(1 - n_S\right) \exp{\left[-1 - N \left(1 - n_S\right)\right]}. 
\end{equation}
Then, for the Planck-allowed values of $1 - n_S < 0.04$ and $N < 60$, we have a lower bound on $r$ of 
\begin{equation}
    r > 0.01. 
\end{equation}
Potentials for which the leading operator is nonrenormalizable, such as 
\begin{equation}
       V\left(\phi\right) = V_0\left[1 -  \left(\frac{\phi}{\mu}\right)^p + \cdots \right], \quad p > 4,
\end{equation}
have a scalar spectral index of
\begin{equation}
    n_S = 1 - \frac{2}{N} \left(\frac{p - 1}{p - 2}\right),
\end{equation}
and are compatible with the Planck bound on $n_S$ for $p \geq 7$,  and there is no lower bound on $r$,
\begin{equation}
    r < 8 \frac{p}{N \left(p - 2\right)} \left[\frac{1}{N p \left(p - 2\right)}\right]^{p / \left(p - 2\right)}.
\end{equation}
Here, compatibility with the data comes with the price of nonrenormalizability, and requires the presence of a symmetry which suppresses all operators of order $p < 5$ to prevent loop corrections from generating these operators. In this paper we consider another option. The bounds (\ref{eq:quadnS}) and (\ref{eq:quadr}) depend on the assumption that the leading behavior is of the form (\ref{eq:quadpotential}) over the \textit{entire} 60 e-folds of inflation. From the standpoint of effective field theory, this is a dubious assumption, since the field excursion in this case is of order $\Delta \phi > M_{\mathrm P}$. It is reasonable to expect that even if the potential is well-approximated by Eq. (\ref{eq:quadpotential}) near $N = 60$, higher-order operators will become significant near the end of inflation. We can parameterize our ignorance of the form of the effective potential during this final stage of inflation by allowing the field value at the end of inflation to be  a free parameter, allowing for arbitrary mapping between the wavenumber $k$ of a perturbation mode and the number of e-folds $N\left(k\right)$ at which it exited the horizon. In this paper, we show that adding this additional free parameter results in a corresponding freedom in the value of the tensor/scalar ratio $r$, such that $r$ can be arbitrarily small, while still satisfying the Planck bound on $n_S$ and allowing for a field excursion $\Delta \phi \ll M_{\mathrm P}$, consistent with the swampland distance conjecture.

\section{Constant roll inflation in the language of the potential}
To discuss the extra degree of freedom in terms of the potential, we begin by considering cosmological inflation driven by a single, minimally coupled canonical scalar field, with field dynamics governed by the Friedmann equation, 
\begin{equation}
    H^2=\frac{1}{3M_{\mathrm P}^2}\left(\frac{1}{2}\dot\phi^2+V(\phi)\right),
\end{equation}
and the equation of motion,
\begin{equation}
    \ddot \phi + 3H\dot\phi+V'(\phi)=0,
\end{equation}
where a prime indicates a derivative with respect to the inflaton field $\phi$, and an overdot indicates a  derivative with respect to coordinate time t.
We assume no spatial curvature and a Friedmann-Robertson-Walker metric of the form 
\begin{equation}
    ds^2=dt^2-a^2(t)d\textbf{x}^2,
\end{equation}
with the Hubble parameter H defined as 
\begin{equation}
    H\equiv \left(\frac{\dot a}{a}\right).
\end{equation} 
If the evolution of the scalar inflaton field $\phi$ is monotonic, we can write the Hamilton-Jacobi equations for the system as
\begin{equation}
    -2M_{\mathrm P}^2H'=\dot\phi
\end{equation}
and 
\begin{equation}
    2\left[H'(\phi)\right]^2-\frac{3}{M_{\mathrm P}^2}H^2(\phi)+\frac{1}{M_{\mathrm P}^4}V(\phi)=0.
    \label{solveforV}
\end{equation}
We can now define the Hubble slow roll parameters in terms of $H(\phi)$:
\begin{equation}
    \epsilon \equiv \frac{\dot \phi^2}{2M_{\mathrm P}^2H^2}=2M_{\mathrm P}^2\left(\frac{H'(\phi)}{H(\phi)}\right)^2,
    \label{epsdef}
\end{equation}
and 
\begin{equation}
    \eta \equiv \frac{-\ddot \phi}{H\dot\phi}=2M_{\mathrm P}^2\frac{H''(\phi)}{H(\phi)}.\label{etadef}
\end{equation}
The parameters $\epsilon$ and $\eta$ are related to the tensor/scalar ratio $r$ and the scalar spectral index $n_S$ by 
\begin{equation}
    r=16\epsilon(\phi_*)
    \label{rdef}
\end{equation}
and 
\begin{equation}
    n_S-1=-4\epsilon(\phi_*)+2\eta(\phi_*).
\end{equation}
Here the observable values calculated from the CMB are set by the value of $\phi_*$, where $\phi_*$ indicates the value of $\phi$ approximately 60 e-folds of expansion before the end of inflation, when the modes cross outside the horizon and freeze out. Note that lower values of $r$ require lower values of $\epsilon_*$, meaning that reducing $\epsilon_*$ enough to achieve $r-$values smaller than $10^{-2}$ or so results in the value of $n_S$ being almost completely set by the value of $\eta$.
For the quadratic potential (\ref{eq:quadpotential}), in the limit $\phi \rightarrow 0$ and $\epsilon \ll \eta$,
\begin{equation}
    \eta \simeq M_{\mathrm P} \frac{V''\left(\phi\right)}{V\left(\phi\right)} \rightarrow \mathrm{const.}
\end{equation}
We are therefore interested in the dynamics of constant roll inflation.

The constant roll inflation model was developed by Motohashi et. al. \cite{Motohashi:2014ppa} as
a generalization of the ultra-slow roll (USR) inflationary solution \cite{Kinney:1997ne,Martin:2012pe,Inoue:2001zt,Kinney:2005vj,Namjoo:2012aa,Huang:2013lda,Mooij:2015yka,Cicciarella:2017nls,Akhshik:2015nfa,Scacco:2015spa,Barenboim:2016mmw,Cai:2016ngx,Odintsov:2017yud,Grain:2017dqa,Odintsov:2017qpp,Bravo:2017wyw,Bravo:2017gct,Dimopoulos:2017ged,Nojiri:2017qvx,Motohashi:2017vdc,Odintsov:2017hbk,Oikonomou:2017xik,Cicciarella:2017nls,Awad:2017ign,Anguelova:2017djf,Salvio:2017oyf,Yi:2017mxs,Cai:2018dkf,Mohammadi:2018oku,Gao:2018tdb,Gao:2018cpp,Anguelova:2018ntr,Mohammadi:2018wfk,Karam:2017rpw}.
This model is characterized by setting $\eta = \mathrm{const.}$, and describes both the inflationary attractor and non-attractor behavior \cite{Morse:2018kda,Lin:2019fcz}. Here we will be interested only in the slow-roll attractor. Taking $\eta = \mathrm{const.}$ and rearranging equation \ref{etadef} to solve for $H(\phi)$ gives
\begin{equation}
    H''(\phi)=\frac{\eta}{2M_{\mathrm P}}H(\phi),
\end{equation}
which has the solution 
\begin{equation}
    H(\phi)=A \cosh{\left[\sqrt{\frac{\eta}{2}}\frac{\phi}{M_{\mathrm P}}\right]} + B \sinh{\left[\sqrt{\frac{\eta}{2}}\frac{\phi}{M_{\mathrm P}}\right]}.
\end{equation}
For the case of a symmetric hilltop about $\phi = 0$, the coefficient of the second term vanishes, leaving us with 
\begin{equation}
     H(\phi)=H_0 \cosh{\left[\sqrt{\frac{\eta}{2}}\frac{\phi}{M_{\mathrm P}}\right]}.
     \label{HCR}
\end{equation}
We next plug this into equation \ref{solveforV} and solve for $V$, finding
\begin{equation}
    V(\phi)=M_{\mathrm P}^2H_0^2 \left(3\cosh^2\left[\sqrt{\frac{\eta}{2}}\frac{\phi}{M_{\mathrm P}}\right]-\eta \sinh^2\left[\sqrt{\frac{\eta}{2}}\frac{\phi}{M_{\mathrm P}}\right]\right).
\end{equation}
Taylor expanding about $\phi = 0$ gives the expansion
\begin{equation}
    \label{eq:CReffective}
     V\left(\phi\right) = M_{\mathrm P}^2 H_0^2 \left\lbrace 3-\left(3-\eta\right)\left[\frac{\eta}{2}\left(\frac{\phi}{M_{\mathrm P}}\right)^2+\frac{\eta^2}{12}\left(\frac{\phi}{M_{\mathrm P}}\right)^4+\cdots\right]\right\rbrace,
\end{equation}
which is quadratic to leading order for $\phi<<M_{\mathrm P},$ corresponding to the generic quadratic hilltop potential \ref{eq:quadpotential}.
Substituting equation \ref{HCR} into equation \ref{epsdef}, we get 
\begin{equation}
    \epsilon(\phi)=\eta \tanh^2\left[\sqrt{\frac{\eta}{2}}\frac{\phi}{M_{\mathrm P}}\right],
    \label{epsphicr}
\end{equation}
so that if we reduce the value of $\phi_*$, we can generate arbitrarily low values of $\epsilon_*$ and thus $r$.

We generate the additional degree of freedom by recognizing that while the effective potential (\ref{eq:CReffective}) is valid near the maximum of the potential where inflation occurs, it is not necessarily valid near the end of inflation, where other operators in general become dynamically relevant. Thus, we cannot in general use the form (\ref{eq:CReffective}) to calculate the end of inflation, and instead we take the field value $\phi_e$ at the end of inflation to be a free parameter. Equivalently, we can treat $\phi_*$  as a free parameter, encoding unspecified late-time behavior. This allows us to tune the mapping between the number of e-folds, $N$, and wavenumber, $k$. As a consequence we can tune the value of $r$ without requiring $n_S$ to substantially shift. In the next section, we use the inflationary flow equations to directly solve for $\epsilon$ as a function of the number of e-folds, $N$, providing an equivalent (and simpler) picture without direct reference to the potential.

\section{Constant roll inflation in the language of flow parameters}
An equivalent way to represent the dynamics is to use the inflationary flow equations \cite{Kinney:2002qn}. In the flow representation, we solve for $\epsilon$ and $\eta$ as functions of $N,$ rather than $\phi$. For constant roll inflation, the infinite system of flow equations reduces to two,
\begin{equation}
\frac{d\epsilon}{dN}=2\epsilon(\eta-\epsilon),
\end{equation}
and
\begin{equation}
\frac{d\eta}{dN}=0,
\end{equation}
since constant roll inflation is characterized by a constant value of $\eta$. Solving this for $\epsilon$, we find that the normalized solution to the flow equations is 
\begin{equation}
    \epsilon(N) = \frac{\epsilon_0 \eta e^{2 \eta N}}{\eta - \epsilon_0 \left(1 - e^{2 \eta N}\right)},
    \label{epsNcr}
\end{equation}
where $\epsilon_0 \equiv \epsilon(N = 0)$, which is a constant of integration defined as the value $\epsilon$ \textit{would} have at the end of inflation in the absence of higher-order operators. This encodes the extra degree of freedom corresponding to the choice of $\phi_e$. Then 
\begin{equation}
    H(N) = \mathcal{H} \sqrt{\frac{\eta - \epsilon_0  \left(1 - e^{2 \eta N}\right)}{\eta - \epsilon_0}},
\end{equation}
where $\mathcal{H} = H(N \rightarrow \infty)$, and
\begin{equation}
    \phi(N) = M_{\mathrm P} \tanh^{-1}\left[\epsilon(N)\right],
\end{equation}
where we have normalized the solution such that $\phi(N \rightarrow \infty) \equiv 0$, as opposed to anchoring $N$ at the end of inflation, $\phi(N=0) \equiv \phi_e,$ with $\epsilon(\phi_e)=1.$  If we use the conventional end condition for inflation of $\epsilon_0=1,$ we can plug \ref{epsNcr} with $N=60$ and $\epsilon_0=1$ into \ref{rdef}, finding the value of the tensor/scalar ratio to be
\begin{equation}
    r=16\frac{\eta e^{120\eta}}{\eta-1-e^{120\eta}}.
    \label{re1}
\end{equation}
Taking $\eta=-0.0175,$ we find $r=0.03$ and $n_s=0.957$, in mild tension with existing data \cite{Planck:2018jri,Easther:2021eje}. This model would be definitively ruled out by even a modest reduction in the upper bound on $r$. If, however, we allow inflation to end with $\epsilon_0 \neq 1,$  we get a more general version of equation \ref{re1}:
\begin{equation}
    r = \frac{16 \epsilon_0 \eta e^{2 \eta N_*}}{\eta - \epsilon_0 \left(1 - e^{2 \eta N_*}\right)}.
\end{equation}
By decreasing the value of the constant of integration, $\epsilon_0,$ we have the freedom to lower $r$ as far as required, for any value of $\eta$ consistent with observation. This is the main result of this paper. We note that this provides a simple counterexample to the argument in Ref. \cite{Easther:2021eje} that a low tensor/scalar ratio should be generically correlated to a large running of the scalar spectral index, using a three-parameter flow expansion. Here, $r$ can be tuned to an arbitrarily low value, while the running remains exactly zero.

\section{Conclusions}
In this work, we show that future reductions in the upper bound on the tensor/scalar ratio, $r$, \cite{CMB-S4:2016ple} cannot rule out hilltop-type single-field inflationary models, provided we allow for an extra degree of freedom corresponding to unknown dynamics at the end of inflation. We construct an exactly solvable realization of this in the constant-roll inflation scenario, for which the second slow roll parameter $\eta$ is exactly constant, and the additional degree of freedom appears as a parametrically adjustable mapping between the number of e-folds $N$ and the wavenumber $k$ of primordial perturbations. This is well-motivated by the swampland distance conjecture, which suggests that an effective field theory expansion for inflation is inconsistent for field excursions $\Delta\phi > M_{\mathrm P}$, since Planck-suppressed nonrenormalizable operators will become significant late in the inflationary epoch. This relevance of this construction for observation is obvious: there is no inherent lower bound on the tensor/scalar ratio, even in ``simple'' canonical single-field inflation models. 

\section*{Acknowledgments}
This work is supported by the National Science Foundation under grants NSF-PHY-1719690 and NSF-PHY-2014021, and by the Indian Institute of Technology, Madras.  This work was performed in part at the University at Buffalo Center for Computational Research. We thank Richard Easther for comments on a draft of this work.

\bibliographystyle{apsrev4-1}
\bibliography{paper}
\end{document}